\documentclass[prl,twocolumn]{revtex4}

\pdfoutput=1

\usepackage[T1]{fontenc}

\usepackage{units}
\usepackage{verbatim}

\expandafter\let\csname equation*\endcsname\relax
\expandafter\let\csname endequation*\endcsname\relax

\usepackage{amsmath}
\usepackage{amssymb}

\usepackage{graphicx}
\usepackage{esint}
\usepackage{pdfpages}

\makeatletter
\@ifundefined{textcolor}{}
{%
	\definecolor{BLACK}{gray}{0}
	\definecolor{WHITE}{gray}{1}
	\definecolor{RED}{rgb}{1,0,0}
	\definecolor{GREEN}{rgb}{0,1,0}
	\definecolor{BLUE}{rgb}{0,0,1}
	\definecolor{CYAN}{cmyk}{1,0,0,0}
	\definecolor{MAGENTA}{cmyk}{0,1,0,0}
	\definecolor{YELLOW}{cmyk}{0,0,1,0}
}

\usepackage{units}
\usepackage{lmodern}
\usepackage{float}

\usepackage{lipsum}
\usepackage{placeins}
\usepackage{dsfont}

\newcommand{\ket}[1]{| #1 \rangle}

\newcommand{\bk}[1]{\left\langle #1 \right\rangle}

\newcommand{\gu}{\gamma_\uparrow}
\newcommand{\gs}{\gamma_{\Sigma}}

\usepackage[english]{babel}

\makeatother

\begin{document}
	
	\title[]{Calorimetric measurement of work for a driven harmonic oscillator}

	\author{Rui Sampaio}
	\affiliation{COMP Center of Excellence, Department of Applied Physics, Aalto 
	University,
		P.O. Box 11000, FI-00076 Aalto, Finland}

	\author{Samu Suomela}
	\affiliation{COMP Center of Excellence, Department of Applied Physics, Aalto 
	University,
		P.O. Box 11000, FI-00076 Aalto, Finland}

	\author{Tapio Ala-Nissila}
	\affiliation{COMP Center of Excellence, Department of Applied Physics, Aalto 
	University,
		P.O. Box 11000, FI-00076 Aalto, Finland}
	\affiliation {Department of Physics, P.O. Box 1843, Brown University, 
	Providence, Rhode Island 02912-1843, U.S.A.}

	\begin{abstract}
		A calorimetric measurement has recently been proposed as a promising 
		technique to measure thermodynamic quantities in a dissipative 
		superconducting qubit. These measurements rely on the fact that the system 
		is projected into energy eigenstates whenever energy is exchanged with the 
		environment. This requirement imposes a restriction on the class of systems that can be measured in this way. 
		Here we extend the calorimetric protocol to the measurement of a driven quantum harmonic oscillator. We employ a scheme based on the two-level approximation to define a new work quantity and show how its statistics relates to the standard two-measurement protocol.
		We find that for the average work the two-level approximation holds in the underdamped regime for short driving times and, in the overdamped regime, for any driving time. However, this approximation fails for the variance and higher moments of work at finite temperatures. Furthermore, we show how to relate the work statistics obtained through this scheme 
		to the work statistics given by the two-measurement protocol.
		
	\end{abstract}
	\maketitle
	
	\section{Introduction}

	Measuring thermodynamic properties of open quantum systems has proven to be a 
	challenging problem. As proposed by Crooks \cite{CrooksJSM2008}, 
	measurement of heat exchanged between the system of interest and its 
	surroundings should be accomplished by measuring the environment only. To this end, many theoretical 
	approaches have been developed in the context of fluctuation theorems 
	%\cite{Esposito:2009aa,Campisi:2011aa,talkner2009fluctuation,Suomela:2015aa,Solinas:2015aa,andrieux2009fluctuation,horowitz2012quantum,
	% chetrite2012quantum, PhysRevE.89.012127, Campisi2009JPA, 
	%PhysRevE.90.022103, 
	%PhysRevE.92.032129,PhysRevE.86.031127, PhysRevE.93.022126, 
	%PhysRevE.89.052116, PhysRevX.5.031038, Prasanna2015NJP,PhysRevE.89.032114, 
	%PhysRevE.88.052121, Chetrite:2010aa} 
	but definitive experimental measurement of open system dynamics have not 
	been attained yet (see for example Ref. \cite{Campisi:2011aa} and references 
	therein). A major obstacle is that current experimental techniques rely on 
	projective measurements of the total system under unitary (closed) evolution 
	\cite{Batalhao2014,An2015}. This leads to an impractical setup for systems 
	coupled to large environments such as the case of small electronic devices.
	
	To address this problem, a measurement scheme has been proposed for a dissipative superconducting qubit based on calorimetry \cite{Pekola2013,Viisanen2015,Pekola:2015aa}. 
	Calorimetric measurements use the concept of \textit{quantum jumps} (QJ) which arise naturally in indirect measurements schemes. 
	When heat is emitted or absorbed by the environment (the calorimeter in this case), its effective temperature changes and the system state collapses (a jump occurs). 
	This energy exchange with the environment can be tracked continuously by monitoring temperature fluctuations in the calorimeter reducing an intractable number of degree of freedom to just one.
	A key point in the calorimetric measurement is that, for the two-level-system (TLS) initially proposed in \cite{Pekola2013}, when heat is exchanged with the environment, the system is projected into an energy eigenstate.
	Therefore, the internal energy change can be inferred from the amount of heat exchanged with the calorimeter. 
	Conversely, the calorimeter protocol cannot be straightforwardly implemented if the system is not 
	projected to an energy eigenstate, because the change in internal energy cannot then be correctly inferred.

	%In reality however, the TLS is only an approximation of a system comprising 
	%of many levels. The possibility of ignoring these extra levels and 
	%approximate the system as having only two states turns out to be crucial in 
	%modeling and measuring low temperature quantum 
	%systems\cite{Clarke:2008aa,Morello2014aa,Small1999aa}.

	\begin{figure}[]
		\centering{} \includegraphics[width=1\columnwidth]{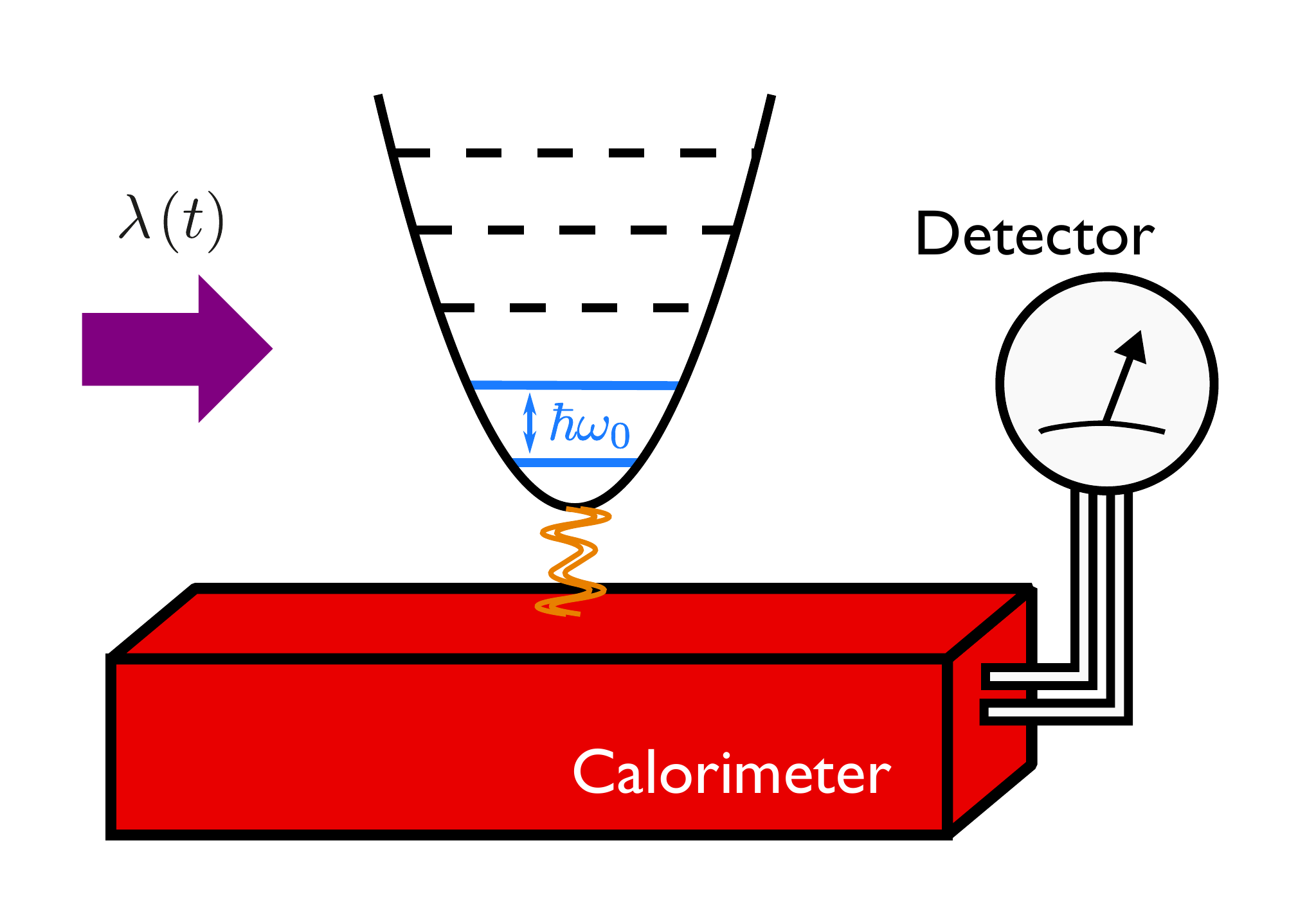}
		\caption{Schematic illustration of the calorimetric measurement setup. The system is a driven QHO coupled to a heat bath (calorimeter). Energy exchange between system and calorimeter is monitored continuously by a detector. Because the system is driven, from the point of view of the measurement of single quanta the system cannot be distinguished from a two-level system. 
		This will lead to errors inferring the internal energy of the system when higher levels become populated.
			\label{fig:intro}}
	\end{figure}

	In the present article, we consider a weakly driven quantum harmonic oscillator (QHO) with equally spaced
	energy levels as our model system. 
	When driven into a coherent energy superposition, the QHO will not jump, in general, to an energy eigenstate. 
	Moreover, if weakly driven, it cannot be distinguished from a TLS from the point of view of single quanta exchange with the calorimeter. 
	To study the applicability of the calorimetric protocol to this case, we define a new work quantity where the change in internal energy is inferred as if the QHO were a two-level system. We then compare the statistics of this new quantity with the standard two-measurement protocol under different driving conditions. 
	
	\section{The model}
	
	The  Hamiltonian is given by $ \hat{H}_{S}(t)=\hat{H}_{0} + \lambda(t)(\hat{a}^{\dagger}+\hat{a}) $, where $ \hat{H}_0 = \hbar\omega_{0}\hat{a}^{\dagger}\hat{a} $ is the standard QHO Hamiltonian,  $\hbar\omega_{0}$ is the level spacing, $\hat{a}^{\dagger}$ is the creation operator, $\hat{a}$ is the annihilation operator, and $\lambda(t)=\lambda_{0}\sin(\omega_{0}t)$ is a resonance, periodic external drive (cf. Fig. \ref{fig:intro}).  
	The discussion will be restricted to weak driving from $t=0$ to $t=T$ such that $\lambda_{0}=0$ for $t<0$ and $t>T$, and we set $\lambda_{0} = 0.01 \hbar\omega_0\ll\hbar\omega_{0}$. We ignore the zero-point-energy contribution $ \hbar\omega_0/2 $. The Hamiltonian can be further simplified by changing to the interaction picture and employing the rotating wave approximation yielding the time-independent Hamiltonian 
	\begin{equation}
	\hat{H}_{S}^{I}=\hat{H}_{0}+\frac{\lambda_{0}}{\sqrt{2}}\hat{P},
	\end{equation}
	where $\hat{P}=i(\hat{a}^{\dagger}-\hat{a})/\sqrt{2}$ is the dimensionless momentum operator and the superscript $I$ denotes the interaction picture with respect to $\hat{H}_0$. 
	The QHO is then coupled to an environment (hereafter referred to as the calorimeter) which is continuously monitored. This causes the evolution to be stochastic. It is particularly relevant from the experimental point of view to formulate the evolution of the system via stochastic trajectories in Hilbert space \cite{breuer_OQS,Molmer1993,Pekola2013}. 
	A single trajectory is described by a sequence of $N$ jumps $\{i_{1},\dots,i_{j},\dots,i_{N}\}$ at times $\{t_{1},\dots,t_{j},\dots,t_{N}\}$, with $t_{1}<\dots<t_{j}<\dots<t_{N}$ together with a non-hermitian evolution operator $\hat{U}_{\mathrm{nh}}(t)$ such that
	\begin{equation}
	|\psi_{\tau}\rangle=\hat{U}_{\mathrm{nh}}(\tau-t_{N})\hat{C}_{i_{N}} 
	\dots\hat{U}_{\mathrm{nh}}(t_{2}-t_{1})\hat{C}_{i_{1}} 
	\hat{U}_{\mathrm{nh}}(t_{1})|\psi_{0}\rangle,\label{eq:traj_njumps}
	\end{equation}
	where $|\psi_{\tau}\rangle$ and $|\psi_{0}\rangle$ denote the states at times $t = \tau$ and $t=0$, respectively. 
	The operators $ \hat{C}_{i_j} $ are called \textit{jump operators} and describe back-action from 
	the calorimeter on the system whenever the energy of the calorimeter changes. 
	The non-hermitian evolution operator $ \hat{U}_\mathrm{nh} $ describes the evolution of the system when the energy of the calorimeter remains constant. 
	Naturally, the squared norm of the wave function is no longer conserved, but instead interpreted as the probability for a particular trajectory to be observed. 
	For the particular experimental setup proposed in Ref. \cite{Pekola2013} the jump operators are
	\begin{align}
	\hat{C}_{0} & =\gamma_{0}^{\nicefrac{1}{2}}\hat{a};\label{eq:C_d}\\
	\hat{C}_{1} & =\gamma_{1}^{\nicefrac{1}{2}}\hat{a}^{\dagger},\label{eq:C_u}
	\end{align}
	where $\gamma_{0} = \gamma [N(\beta) + 1]$ and $\gamma_{1} = \gamma N(\beta)$ are the relaxation rates corresponding to heat absorption and emission by the bath, respectively, $ \gamma $ is the coupling strength between the 
	calorimeter and the system, and $ N(\beta) = [\exp(\beta\hbar\omega_0)-1]^{-1} $ is the average occupation number. 
	The non-hermitian evolution operator $\hat{U}_{\mathrm{nh}}(t)$ is given explicitly by
	\begin{equation}
	\hat{U}_{\mathrm{nh}}(t)=\exp\left[-\frac{i}{\hbar}\left(\frac{\lambda_{0}}{\sqrt{2}}\hat{P}+\hat{D}\right)t\right],
	\label{eq:U_nh_ho}
	\end{equation}
	where $\hat{D} = -i\hbar/2\left(\gamma_{\Sigma}\hat{a}^{\dagger}\hat{a}+\gamma_{1}\right)$, with $\gamma_{\Sigma}=\gamma_{0}+\gamma_{1}$.
	
	Finally, we note this formulation is equivalent to the Lindblad master equation and therefore applies only in the weak coupling limit $ \hbar\gamma_0 \ll \hbar\omega_0 $, $k_BT \gg \hbar\gamma $ \cite{breuer_OQS}.

	\subsection{Calorimetric work}	
	
	In the spirit of stochastic thermodynamics \cite{Seifert2012RPP}, work is defined as a stochastic variable, $W$, in accordance with the first law of thermodynamics
	\begin{equation}
	W=\Delta U+Q,
	\label{eq:W=DU+Q}
	\end{equation}
	where $\Delta U$ is the change in the internal energy of the system and $Q$ is the heat exchanged with the environment. The latter is what is measured directly by the calorimetric protocol, related to the jumps in the trajectory as discussed above. Since each jump is associated with a well-defined energy change of the calorimeter ($\pm\hbar\omega_{0}$), the total heat exchanged in a trajectory of $ N $ jumps is given by 
	\begin{equation}
	Q=\hbar\omega_{0}\sum_{j=1}^{N}(-1)^{i_{j}} .
	\label{eq:Q}
	\end{equation}

	The internal energy change is defined through the two measurement protocol, $ \Delta U = E_m - E_n $, where $ E_n $ and $ E_m $ are the result of projective energy measurements performed on the system at beginning ($ t = 0 $) and end ($ t = T $) of the driving protocol, respectively. We shall refer to work as defined by Eq. (\ref{eq:W=DU+Q}) as the \textit{projective work}. 
	For a TLS, the change in internal energy is known \emph{exactly} in the calorimetric measurements from the \emph{guardian photons} -- the last photon exchanged before the driving starts and the first photon exchanged after the driving ends. 
	This comes from the fact that the two-level system always jumps to an energy eigenstate (this is easy to see by applying either operators in Eqs. (\ref{eq:C_d}) or (\ref{eq:C_u}) to an arbitrary state $|\psi\rangle = c_0|0\rangle+c_1|1\rangle$). 
	
	For the QHO, if we consider a general state $\ket{\psi}=\sum_{n}c_{n}\ket{n}$ it is then clear that the action of any of the operators in Eqs. (\ref{eq:C_d}) and (\ref{eq:C_u}) will not, in general, project the system into an eigenstate of $\hat{H}_{0}$. This implies that $\Delta U$ cannot be inferred \emph{exactly} from the guardian photons.\footnote{ Strictly speaking, $\Delta U$ cannot even be attributed a well defined value since the oscillator will in general be left at an energy superposition. However, we can follow a statistical interpretation and interpret the results in terms of averages as used in other works in connection to thermodynamic cycles \cite{PhysRevE.93.022122,Ronnie2014,Dong:2015aa,PhysRevLett.112.030602,Perarnau-Llobet:2015aa}.}
	
	We now invoke the TLA to infer the change in internal energy as if the QHO were a TLS. Let $ \ell_i \in \{0,1\} $ represent the first guardian photon and $ \ell_f \in \{0,1\} $ the last guardian photon. Then the change in internal energy attributed to each trajectory is defined as $ \Delta U_c = \ell_f - \ell_i $, and the work is given by
	\begin{equation}
	W_{c}= \Delta U_c + Q\,.
	\label{eq:W_c-1}
	\end{equation}
	To make a clear distinction between $ W $ and $ W_c $, we shall refer to $ W_c $ as the \emph{calorimetric work} for brevity, but its precise definition is \emph{work measured in the calorimetric protocol under the two level approximation}. Note that this TLA is reflected on the value attributed to the internal energy change. 
	It does \textit{not} relate to the more familiar two level approximation of the dynamics commonly used in optical, atomic or cavity QED physics. 
	In all of the following results we always consider the time evolution of the driven, damped QHO.

	\begin{figure}[]
		\centering{} \includegraphics[width=1\columnwidth]{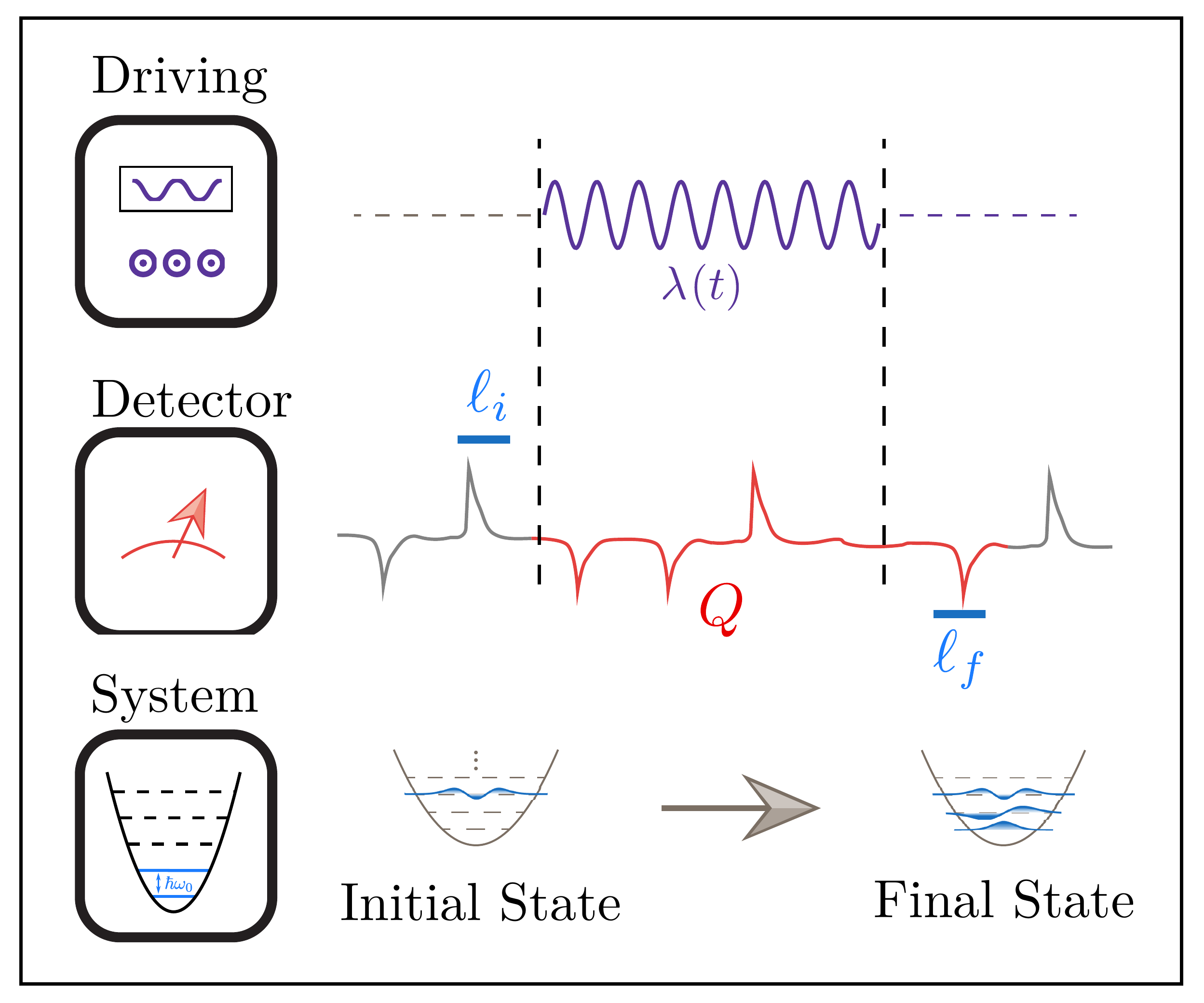}
		\caption{Illustration of a single trajectory of the driven QHO. The vertical lines indicate the beginning and end of the drive. The detector sees a series of clicks corresponding to emission or absorption of energy as if the system were a TLS. However, the system can start from any energy eigenstate of the QHO and will, in general, end at a superposition in the end. $ \ell_{i} $ and $ \ell_{f} $ denote the guardian photons used to infer the change in internal energy (see text for details).
		}
		\label{fig:traj}
		
	\end{figure}

	\section{Results}
		
	We are interested in comparing the statistics of $W_{c}$ to that of $W$, in particular the distribution's average and variance as a function of the driving time. From Eq. (\ref{eq:W_c-1}) the $ k^{th} $ moment of the calorimetric work is written formally as
	\begin{equation}
	\left\langle W_{c}^{k}(t)\right\rangle  =\sum_{\text{traj}}p_{c}^{\text{traj}}(t) [\hbar\omega_{0}(\ell_{f}-\ell_{i})+Q ]^{k},
	\label{eq:<W_c^k>}
	\end{equation}
	where $\sum_{\text{\ensuremath{\left\{  \text{traj}\right\} } }}$ is shorthand notation for the summation over all possible trajectories. This is weighted by the trajectories' probabilities
	\begin{align}
		\sum_{\text{traj}} & p_{c}^{\text{traj}}(t)\equiv   \sum_{n,m=0}^{\infty}p_{\mathrm{eq}}(n)\sum_{\ell_{i},\ell_{f}=0}^{1}p_{i}(\ell_{i}|n)p_{f}(\ell_{f}|m) \nonumber \\
		                   & \times\sum_{N=0}^{\infty}\sum_{i_{N}=0}^{1}\dots\sum_{i_{1}=0}^{1}\int_{0}^{t}dt_{N}\dots\int_{0}^{t2}dt_{1} \nonumber                           \\
		                   & \times T_N(m,t;i_{N},t_{N};\dots;i_{1},t_{1}|n),
	\label{eq:sum_traj_p_traj^c}
	\end{align}
	where 
	\begin{multline}
	T_N(m,t;i_{N},t_{N};\dots;i_{1},t_{1}|n)= \\
	\left|\left\langle  m\left|\hat{U}_{\mathrm{nh}}(t-t_{N})\hat{C}_{i_{N}}\dots\hat{C}_{i_{1}}\hat{U}_{\mathrm{nh}}(t_{1})\right|n\right\rangle \right|^{2}
	\label{eq:T(UCU)}
	\end{multline}
	is a transmission coefficient encoding the probability for a particular trajectory, $p_{i}(\ell_{i}|n)$ is the probability of having observed $\ell_{i}$ given that the state is initially $|n\rangle$, $p_{f}(\ell_{f}|m)$ is the probability of observing $\ell_{f}$ given that the state after the driving is $|m\rangle$, and we assume that the system starts from thermal equilibrium with $p_{\mathrm{eq}}(n)=(1-e^{-\beta\hbar\omega_{0}})\exp(-\beta\hbar\omega_{0}n)$. 
	
	The main quantities to be evaluated are the guardian photons probabilities' $p_{i}(\ell_{i}|n)$ and $p_{f}(\ell_{f}|\psi_{T})$, and the transmission coefficient $T(m,t;i_{N},t_{N};\dots;i_{1},t_{1}|n)$. The former are easily evaluated and given by
	\begin{align}
	p_{f}(0|m) & = \frac{\gamma_{0}m}{\gamma_{0}m+\gamma_{1}(m+1)}; \label{eq:pf_0} \\
	p_{f}(1|m) & = \frac{\gamma_{1}(m+1)}{\gamma_{0}m+\gamma_{1}(m+1)}, \label{eq:pf_1}
	\end{align}
	and
	\begin{align}
	p_{i}(0|n) & =  \frac{\gamma_{1}(n+1)}{\gamma_{1}(n+1)+\gamma_{0}n};\label{eq:p_i_0}\\
	p_{i}(1|n) & =  \frac{\gamma_{0}n}{\gamma_{1}(n+1)+\gamma_{0}n}. \label{eq:p_i_1}
	\end{align}
	For the transmission coefficient $T_N$, an analytical solution for an arbitrary number of jumps $ N $ is, in general, not available for practical purposes. However, we can treat it perturbatively in the limit $ \hbar\gs \ll \lambda_0 $ by expanding the evolution operator up to second order in $ \gs/2 $.
	
	We next present analytic results in this limit for two cases: First, the case where the driving time is much shorter than the thermal relaxation such that the evolution operator can be approximated as unitary and there are no jumps in the trajectories, and second, where corrections up to one jump per trajectory are included.

	\subsection{Underdamped regime}
	\subsubsection{Unitary limit}
		
	If the driving period $T$ is short enough such that $1/T\gg\gamma_{\Sigma}$, the second term inside the exponential in Eq. (\ref{eq:U_nh_ho}) can be dropped and the evolution is approximated as unitary 
	\begin{equation}
	\hat{U}_{\mathrm{nh}}(t)\approx\exp \left( -\frac{i}{\hbar}\frac{\lambda_{0}t}{\sqrt{2}}\hat{P} \right) \equiv\hat{U}_{\mathrm{u}}(t)\quad,
	\label{eq:U_D}
	\end{equation}
	where the subscript $ \mathrm{u} $ is used to denote unitary dynamics. For this case there is no heat exchange during the driving period and we only have to account for the guardian photons, so we write $W_{c}=\hbar\omega_{0}[\ell_{f}-\ell_{i}+(-1)^{\ell_{f}}]$, where the last term is the heat contribution from the last guardian photon. Taking into account only the no-jump term in Eq. (\ref{eq:sum_traj_p_traj^c}) and assuming that the only the two lowest levels are relevant in thermal equilibrium, we can write
	\begin{equation}
	\bk{ W_c^k (t) }_{\mathrm{u}}\approx (\hbar\omega_0)^k\frac{w_{0k} + w_{1k}e^{-\beta\hbar\omega_0}}{1+e^{-\beta\hbar\omega_{0}}},
	\label{eq:<W_c^k>_u}
	\end{equation}
	where the coefficients 
	\begin{multline}
	w_{nk} = \sum_m\sum_{\ell_{i},\ell_{f}} p_{i}(\ell_{i}|n)  p_{f}(\ell_{f}|m) \\  \times T_0(m,t|n) [\ell_{f}-\ell_{i}+(-1)^{\ell_{f}}]^{k}
	\label{eq:w_{nk}}
	\end{multline}
	can be expressed in terms of hypergeometric functions (see the supplementary material). 
	For projective work the average and variance are easily evaluated from Eq. (\ref{eq:W=DU+Q}) for any initial temperature. They are given by (see the supplementary material)
	\begin{align}
	\left\langle W(t)\right\rangle _{\mathrm{u}} & =\hbar\omega_{0}\frac{\lambda_{0}^{2}t^{2}}{4}\equiv\hbar\omega_{0}\mu(t)\label{eq:avg_W_u} ;         \\
	(\sigma_{W}^{2}(t))_{\mathrm{u}}             & =2(\hbar\omega_{0})^{2}[N(\beta\hbar\omega_{0})+\frac{1}{2}] \mu(t)\nonumber                               \\
	& =2\hbar\omega_{0}[N(\beta\hbar\omega_{0})+\frac{1}{2}] \left\langle W(t)\right\rangle_{\mathrm{u}}. 
	\label{eq:var_W_u}
	\end{align}

	To compare $\left\langle W^{k}\right\rangle _{\mathrm{u}} $ and $\left\langle W_{c}^{k}\right\rangle _{\mathrm{u}}$ from Eqs. (\ref{eq:<W_c^k>_u}), (\ref{eq:avg_W_u}) and (\ref{eq:var_W_u}) we start by looking at the zero temperature limit. 
	Taking only the zeroth order term in Eq. (\ref{eq:<W_c^k>_u}) and using $ \beta \rightarrow \infty $, yields
	\begin{align}
	\left\langle W(t)\right\rangle _{\mathrm{u}}     & =\hbar\omega_{0}\mu(t)   ;                                                 \\
	(\sigma_{W}^{2})_{\mathrm{u}}(t)                 & =\left(\hbar\omega_{0}\right)^{2}\mu(t)    ;                               \\
	\left\langle W_{c}(t)\right\rangle _{\mathrm{u}} & =\hbar\omega_{0} [1-e^{-\mu(t)} ] \, ;                             \\
	\left(\sigma_{W_{c}}^{2}\right)_{\mathrm{u}}(t)  & =(\hbar\omega_{0})^{2} e^{-2\mu(t)} [e^{\mu(t)}-1]\,.
	\end{align}
	We first note that the projective work statistics can be easily regained by measuring the calorimetric work, even when the TLA is clearly violated. 
	One particular aspect of this relation is that it does not depend on the details of the driving or bath coupling and we can envision a case where these parameters can be extracted from calorimetric measurements without the need to perform projective measurements at all. 
	Second, if we look at the short time behavior of the average and the variance we see that the calorimetric work reproduces the projective work to first order, therefore providing a very good approximation for short driving periods. 
	This is an expected result since for short driving periods the state of the system is only weakly perturbed from its equilibrium state, where the TLA is justified. In the long drive time limit, the calorimetric work average and variance asymptotically approach $\hbar\omega_{0}$ and zero, respectively.
	This can be explained by looking at Eqs. (\ref{eq:pf_0}) and (\ref{eq:pf_1}) in the zero temperature limit where $ \gamma_1 \to 0 $. The only two possibilities are that a photon will be emitted to the bath or no photon will be observed. The probability for the latter is proportional the probability of finding the system in the ground state after the driving. In the limit $ T \to \infty $, this probability goes to zero. Thus, the last guardian photon will be observed from the system to the bath with probability one which means $ W_c = 1 $ for all trajectories (since the system always starts from the ground state).
	
\begin{figure}
		\includegraphics[width=1\columnwidth]{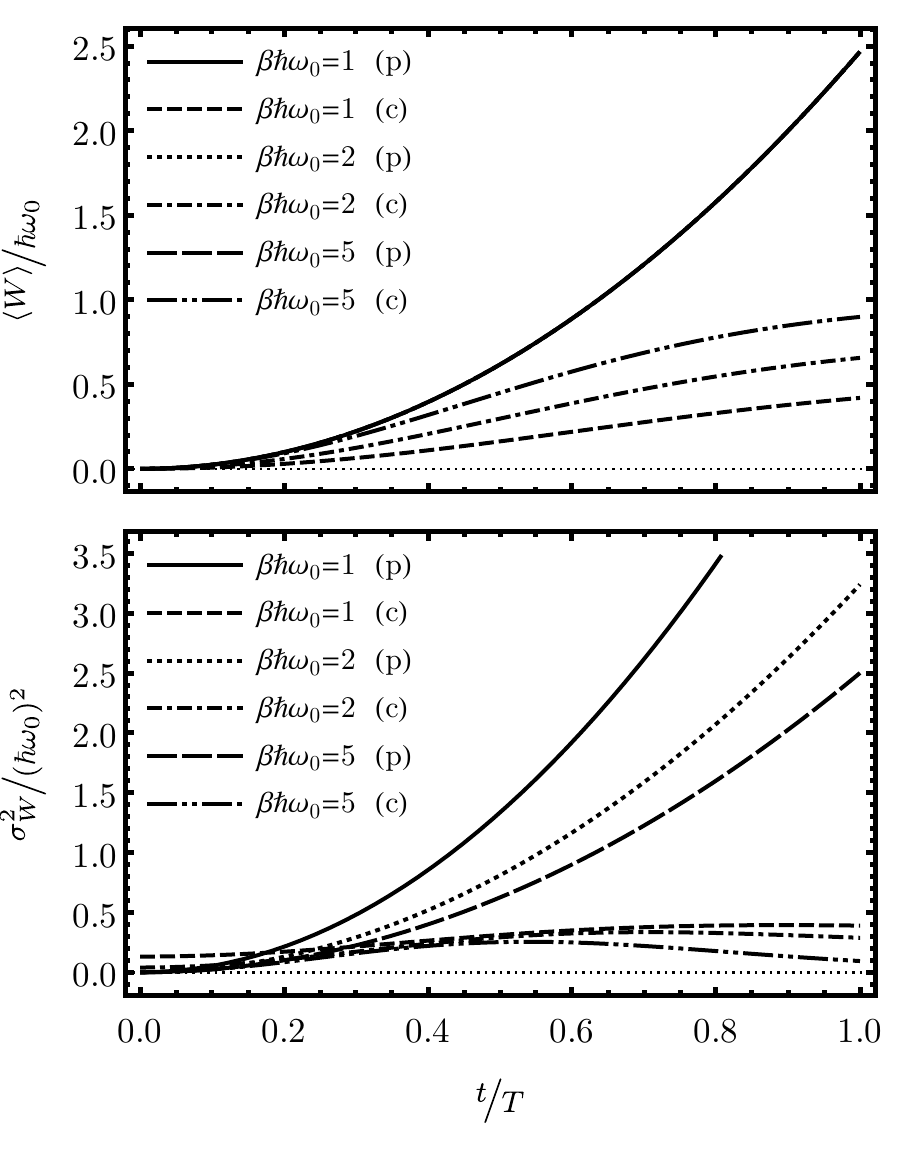}
		\caption{Temperature dependence of the average and variance for calorimetric and projective work in the unitary limit. The letter "p" denotes projective work and the letter "c" denotes calorimetric work. The figure shows data for three different temperatures $ \beta\hbar\omega_0 = 1, 2 $ and $ 5 $. The parameters used are $\lambda_{0}=0.01\hbar\omega_0$, $ \hbar\gamma=0.01\lambda_{0} $, and $T=\hbar\pi/\lambda_{0}$.
			\label{fig:avg_var_temp}}
	\end{figure}

	A similar analysis holds at a finite temperature. Figure \ref{fig:avg_var_temp} shows the average and variance of calorimetric and projective work for three different temperatures $ \beta\hbar\omega_0 = 1, 2 $ and $ 5 $. The driving period is $ T = \hbar\pi/\lambda_0 $. For the average, as the temperature increases the calorimetric work decreases, while the projective work remains invariant. For the variance there is an increase with temperature for projective work as expected, but a more subtle behavior for the calorimetric work.
	When temperature is increased we see a non-zero variance as $ t \rightarrow 0 $. Looking at Eqs. (\ref{eq:pf_0})-(\ref{eq:p_i_1}), in the high temperature limit where higher levels are occupied we can approximate $ p_i(\ell_{i}|n) \approx p_f(\ell_{f}|m) \approx 1/2 $. Consequently, the calorimetric moments in Eq. (\ref{eq:<W_c^k>}) are reduced to $ \langle W_c^k \rangle = (\hbar\omega_0)^k/4 \sum_{\ell_{i},\ell_{f}}[ \ell_{f} - \ell_{i} + (-1)^{\ell_{f}}]^k $ at $ t = 0 $. This yields zero for the average and 0.5$ (\hbar\omega_0)^2 $ for the variance. 
	This non-zero variance at $ t = 0 $ is an artifact coming from the wrong inference of the internal energy from the guardian photons. Suppose, for example, that the system started from the first excited state and that the observed initial guardian photon was $ \ell_{i} = 0 $. Because the system is not in the ground state, it is possible to observe the final guardian photon with $ \ell_{f} = \ell_{i} = 0 $. Thus the inferred internal energy change is $ \ell_{f} - \ell_{i} = 0 $, which means $ W_c = 1 $. Therefore, it is possible to have trajectories with $ W_c \neq 0 $ leading to non-zero variance. 
	Naturally, if the system is thermalized and incoherent the change in its internal energy can be correctly inferred by keeping track of all the jumps.

	\subsubsection{Single jump corrections}
	
	\begin{figure}
		\centering
		\includegraphics[width=1\linewidth]{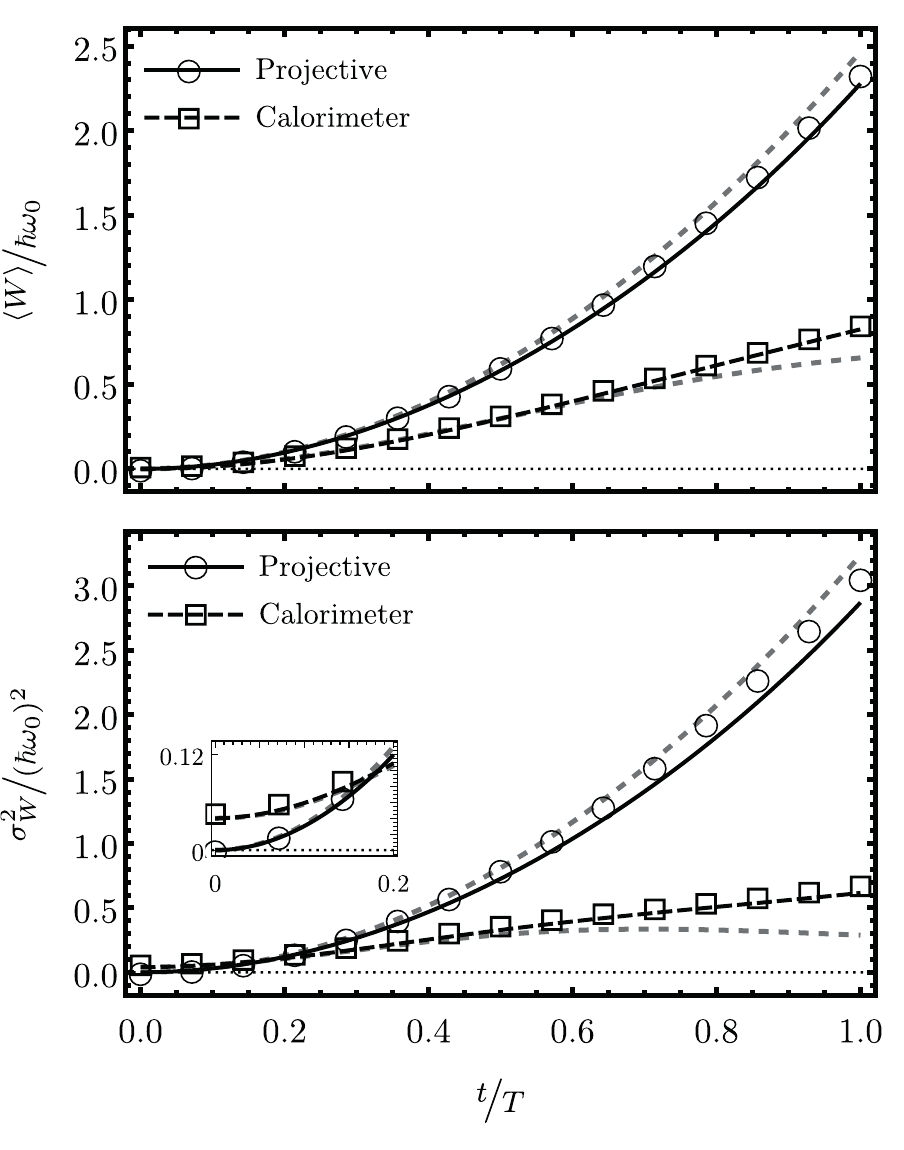}
		\caption{Lowest order corrections to the unitary regime for the average and variance at a fixed temperature ($ \beta\hbar\omega_0 = 2 $). The letter "p" denotes projective work and the letter "c" denotes calorimetric work. The dashed lines show the unitary limit results from Eqs. (\ref{eq:<W_c^k>_u}), (\ref{eq:avg_W_u}) and (\ref{eq:var_W_u}), and the solid lines shown the analytic results with corrections up to two jumps using the perturbative method described in the text. Markers show numerical results using the full evolution operator in Eq. (\ref{eq:U_nh_ho}). The inset shows short-time behavior for $ t/T \in [0,0.2] $. }
		\label{fig:gamma=0.1lmb}
	\end{figure}
		
	If the driving time is comparable to the dissipation rate, $1/T \sim \gamma_{\Sigma}$, jumps must be taken into account. To this end, we treat the dissipation perturbatively with respect to the driving by expanding the evolution operator of Eq. (\ref{eq:U_nh_ho}) up to second order in $\gs/2$:
	\begin{align}
	\hat{U}_{\mathrm{nh}}(t)\approx & e^{-\frac{i}{\hbar}\frac{\lambda_{0}t}{\sqrt{2}}\hat{P}} [1  -\frac{i}{\hbar}\int_{0}^{t}\mathrm{d}t_{1}\hat{D}(t_{1})  \nonumber \\ 
	&  -\frac{1}{\hbar^{2}}\int_{0}^{t}\mathrm{d}t_{1}\int_{0}^{t_{1}}\mathrm{d}t_{2}\hat{D}(t_{1})\hat{D}(t_{2}) ]
	\label{eq:U_nh_expan}
	\end{align}
	where $\hat{D}(t)=-i\gamma_{\Sigma}/2[\hat{H}'(t)+x_0(t)\hat{X}]$, $ x_0(t) = \sqrt{2\mu(t)} $, $\hat{H}'(t)=\hat{a}^{\dagger}\hat{a}+\gu/\gs+\mu(t)$, 
	and $\hat{X}=(\hat{a}^{\dagger}+\hat{a})/\sqrt{2}$. 
	The central quantity to evaluate will be
	\begin{equation}
	u(m,t|n) = U_{0}^{mn}(t) + \frac{\gs}{2}U_{1}^{mn}(t) + \frac{\gs^2}{4}U_{2}^{mn}(t),
	\label{eq:t0}
	\end{equation}
	where $ U_{N}^{mn}(t) $ is the $ N^{th} $ order term of the expansion of $ \langle m | \hat{U}_{\mathrm{nh}} (t) | n \rangle $, as defined in Eq. (\ref{eq:U_nh_expan}) (see the supplementary material). Correction terms for any number of jumps can now be written through Eq. (\ref{eq:t0}).
	 In particular, the transmission coefficient for trajectories with no jumps is given by
	\begin{equation}
	T_0(m,t|n)= \left| u (m,t|n)\right|^2,
	\label{eq:T_0}
	\end{equation}
	and for one jump
		\begin{multline}
		T_{1}(m,t;i_{1},t_1|n) = \gamma_{i_{1}} \vert b_{i_{1}}(t_{1})u(m,t|n) +  \\
		 a_{i_{1}}(t_{1}) \sqrt{n+\delta_{i_1,1}}u(m,t|n+(-1)^{i_1+1}) \vert ^2,
		\label{eq:T_1}
		\end{multline}
	with $a_{i_1}(t)=\exp[(-1)^{i+1}\gamma_{\Sigma}t / 2 ]$ and $b_{i_1}(t) = \lambda_{0}[a_{i}(t)-1]/\hbar\gamma_{\Sigma}$.
	Following this scheme we can calculate the transmission coefficient for two or more jumps but the expressions become increasingly cumbersome with no added physical insight.
	In practice, to evaluate the work moments from Eq. (\ref{eq:<W_c^k>}), the summation over $ m $, $ n $ and $ N $ has to be truncated at some reasonable values depending on the driving time and system parameters. 
	As an example, Fig. \ref{fig:gamma=0.1lmb} shows the deviation from the unitary case (dashed lines in the figure) by considering a coupling strength to the environment $ \hbar\gamma = 0.1 \lambda_0 $, with $ \lambda_0 = 0.01 \hbar\omega_0 $, at 
	fixed temperature ($ \beta\hbar\omega_0 = 2 $). Analytical results (solid lines) are evaluated by considering only the first two levels at the beginning of the protocol ($ n \leq 1 $) and up to the tenth level at the end ($ m \leq 10 $), 
	and trajectories up to two jumps ($ N \leq 2 $). 
	As expected, the deviation is more pronounced the longer the system is driven but the error between the two protocols will decrease as compared to the limit of unitary dynamics. As explained below we expect that the stronger the dissipation as compared to the driving strength, the smaller the error. Moreover, the bounds for the calorimetric average and variance will change. Strictly speaking they are no longer bounded from above since each trajectory can contain any number of jumps. However, for a fixed driving time we expect to see some asymptotic behavior. 
	%since the number of jumps per trajectory decays exponentially beyond the average. 
	
	To numerically validate the approximations made we employed the quantum jump (QJ) method \cite{Molmer1993}  using a ten level system to simulate the QHO \footnote{We numerically verified that there is no change in the moments of work when using more than ten levels for the parameters used here, namely, the driving time and temperature. Also, agreement with the analytic results in the limit of unitary evolution shows that this is a good approximation.}. We used the Monte Carlo Solver 
	from the Quantum Toolbox in Python (QuTiP) \cite{Johansson20121760,Johansson20131234}  for $ 10^5 $ trajectories using the full non-hermitian evolution operator of Eq. (\ref{eq:U_nh_ho}). 
	To calculate the projective work for each trajectory, the measurement process is simulated by drawing a random energy outcome ($ E_m $) weighted by the system state at a given time (say $ t = \tau $) subtracted to the energy outcome at time zero ($ E_n $). 
	The heat is given by Eq. (\ref{eq:Q}) by considering all the jumps up to $ t = \tau $. Calorimetric work is evaluated similarly, with the difference that the system state at $ t = \tau $ is used to evaluate the probabilities of observing a given jump if the driving had stop at that point, given by Eqs. (\ref{eq:pf_0}) and (\ref{eq:pf_1}). 
	Then a random energy outcome is drawn ($ \ell_{f} $) weighted by these probabilities. 
	The same procedure is employed for the first guardian photon ($ \ell_{i} $), using Eqs. (\ref{eq:p_i_0}) and (\ref{eq:p_i_1}). The heat is calculated as in the projective case with the addition of the last guardian photon contribution.
	As can be seen from the Fig. \ref{fig:gamma=0.1lmb} there is a good quantitative agreement between analytic (solid lines) and numerical results (markers). The deviation as time increases is attributed to trajectories containing more than two jumps and higher levels of the system being populated. 	
	
	\subsection{Overdamped regime}	
		
\begin{figure}
		\centering
		\includegraphics[width=1\linewidth]{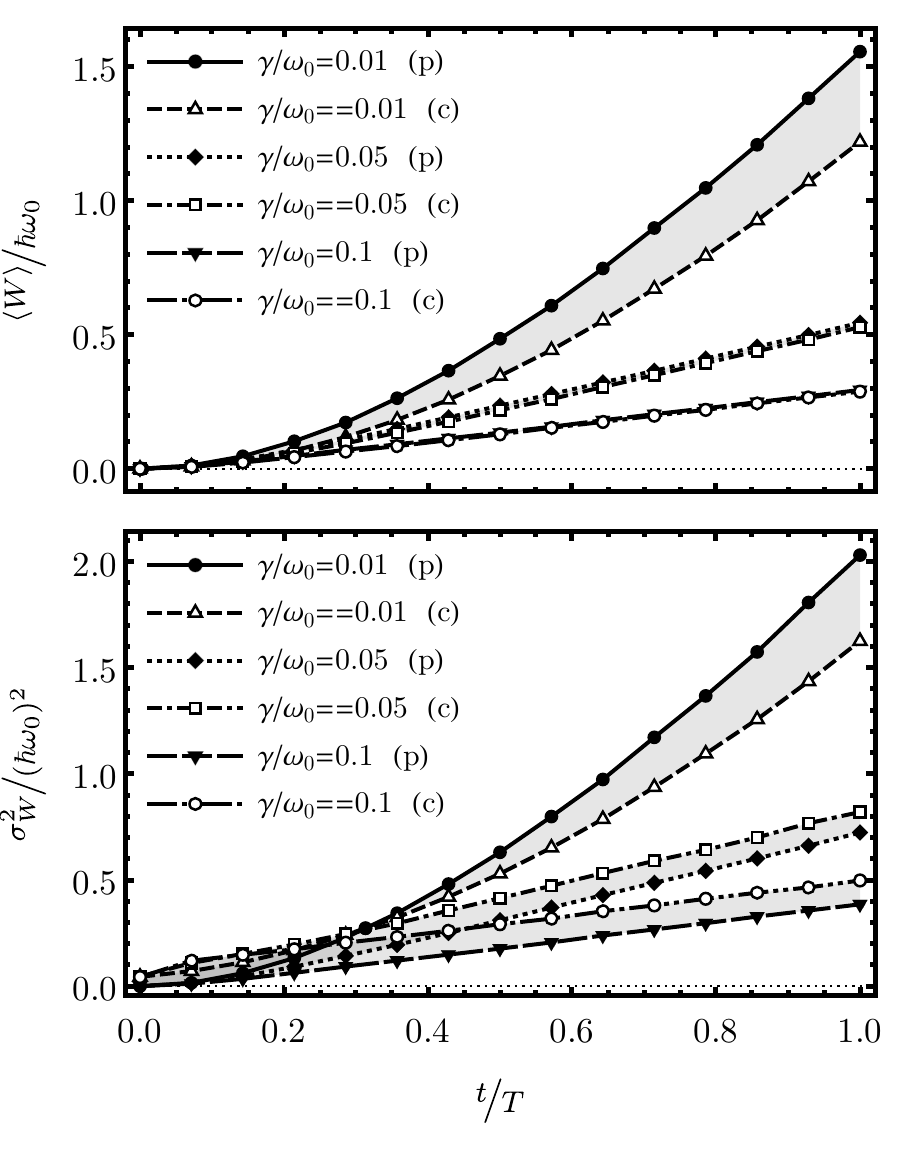}
		\caption{Numerical results for the average and variance of calorimetric and projective work for three different couplings $\hbar\gamma/\hbar\omega_0 =  .01, .05 $ and  $ .1 $ with $ \lambda_0 / \hbar\omega_0 = 0.01 $ and $ \beta\hbar\omega_0 = 2 $. The letter "p" denotes projective work and the letter "c" denotes calorimetric work.}
		\label{fig:strongDissipation}
	\end{figure}

	In the limit $ \hbar\gamma \gg \lambda_0 $ the perturbative method fails and a general analytic solution is not available due to the complicated form of the transmission coefficient in Eq. (\ref{eq:T(UCU)}). 
	Notice that we are still in the the weak coupling limit such that $ \omega_0 \gg \gamma $ holds. 
	It is easy to see that the average calorimetric work will reproduce the average projective work since the relaxation time to equilibrium of the system is much faster than the time required for the driving to push the system to higher levels, and the system remains close to its equilibrium state with a negligible change in its internal energy $U$. Therefore, all the (average) work is dissipated as heat into the reservoir, i.e., $ \bk{W} \approx \bk{Q} $. Since the calorimetric protocol measures the dissipated heat exactly, as the coupling to the environment increases (or the driving strength decreases), the calorimetric average work will reproduce the projective average work. This is, however, not true for the variance, and higher moments in general. 
	Taking the second moment of calorimetric work gives
	\begin{equation}
	\bk{W_c^2} = \bk{\Delta U_c^2} +\bk{Q^2} + 2\bk{\Delta U_c\ Q}.
	\end{equation}
	From the discussion above, the first term has a negligible contribution but the last term will, in general, contribute to the result. Figure \ref{fig:strongDissipation} shows numerical results for the average and variance using the QJ method. As before, we use the Monte Carlo solver from QuTiP for $ 10^5 $ trajectories using the full non-hermitian evolution operator in Eq. (\ref{eq:U_nh_ho}). 
	%For this example we have used fixed temperature $ \beta\hbar\omega_0 = 2 $ and $ \lambda_0 = 0.01 \hbar\omega_0 $ and different couplings to the environment $ \hbar\gamma/\lambda_0 \in \{1,5,10\} $. 
	As it is clear the average calorimetric and projective work will quickly converge to the same limit for $ \hbar\gamma \gg \lambda_0 $. The variance, however will not.

	\section{Summary and conclusions}
	\label{sec:Conclusions}
			
	The recently proposed calorimetric measurement technique presents a promising setup to evaluate thermodynamics quantities in open quantum systems. 
	Here we have looked at the validity of this protocol for a simple model, namely a weakly driven QHO under the assumption that it can be approximated as a two-level system. We have shown how this assumption influences the work measured 
	in the calorimetric protocol and compared to that obtained from idealized projective measurements. In particular, we have shown that the two-level approximation holds for short driving periods and that there is a simple relation between calorimetric and projective work at zero temperature independent of the driving strength and dissipative coupling (within the weak driving assumption $ \hbar\omega_0 \gg \lambda_0 $ and weak coupling $ \hbar\gamma \ll \hbar\omega_0  $ to the environment).  Furthermore, the two-level approximation introduces certain artifacts in the internal energy inference such as a non-zero variance with no driving present. 
	Note that no approximation is needed if the system thermalized and decohered and no driving is present. By keeping track of all the jumps the internal energy change can be inferred precisely. This suggests that each jump carries information that can be use to better infer the internal state of the driven system. 
	If there's strong dissipation compared to the driving strength, the change in the internal energy is negligible and the average calorimetric work reproduced the average projective work.  It should be noted here that technical details regarding calorimetric measurements -- such as finite heat capacity \cite{Pekola2016,PhysRevE.93.062106,Suomela2016b,PhysRevE.94.022123} or incomplete measurement \cite{Viisanen2015} -- are not considered in this work. We expect that the general conclusions drawn here will not change although quantitative changes will certainly appear.
	
	Finally, this work shows that under certain conditions it is possible to infer the correct distribution of thermodynamic quantities from indirect, and possibly incomplete measurements provided the dynamics of the system under investigation is known. This approach can then be extended to other measurements techniques which have to rely on indirect observation of the system's internal state.

	\section*{Acknowledgments}
	
	The authors acknowledge fruitful discussions with Rebecca Schmidt and Jukka Pekola. This work was support by the Academy of Finland through its Centers of Excellence Programme (2015-2017) under project numbers 251748 and 284621. S.S. acknowledges financial support from V\"{a}is\"{a}l\"{a} foundation. The numerical calculations were performed using computer resources of the Aalto University School of Science "Science-IT" project.

	\bibliography{QHO-cal}

\end{document}